\documentclass[journal]{IEEEtran}
\AtBeginDocument{%
  }

\usepackage{adjustbox}

\usepackage{booktabs}

\usepackage{listings}

\usepackage[scaled=.7]{beramono}

\usepackage[T1]{fontenc}

\usepackage{inconsolata}

\usepackage{xcolor}

\usepackage{amssymb}

\usepackage{pifont}

\usepackage[most]{tcolorbox}

\lstdefinestyle{promptstyle}{
  backgroundcolor=\color{gray!3},
  basicstyle=\ttfamily\scriptsize,
  frame=single,
  rulecolor=\color{gray!10},
  breaklines=true
  framerule=0pt,
  xleftmargin=2pt,
  xrightmargin=2pt,
  aboveskip=7pt,
  belowskip=0pt,
  captionpos=b,
  showstringspaces=false,
  morekeywords={Role,Input,Instructions,Output}
}
\newcommand{\inlinecode}[1]{%
  \setlength{\fboxsep}{0.5pt}%
  \colorbox{gray!10}{\footnotesize\texttt{#1}}%
}
\begin{document}

%
\title{Understanding and Mitigating Library-Related Issues in LLM-Generated Code}
%
%
%

\author{
Yacine Majdoub, Rinad Hamid, Eya Ben Charrada, Ahmad Abdellatif, and Haifa Touati
\IEEEcompsocitemizethanks{\IEEEcompsocthanksitem Y. Majdoub, E. Ben Charrada, and H. Touati are with the IReSCoMath Research Laboratory, Faculty of Sciences, University of Gabes, Tunisia . \protect\\ E-mails: \{ yacine.majdoub@enig.rnu.tn, eya.bencharrada@fsg.rnu.tn, haifa.touati@univgb.tn \} }
\IEEEcompsocitemizethanks{\IEEEcompsocthanksitem R. Hamid and A. Abdellatif are with the Department of Electrical \& Software Engineering, University of Calgary, Calgary, AB, Canada.\protect\\ E-mails: \{ rinad.hamid@ucalgary.ca, ahmad.abdellatif@ucalgary.ca \}}
\IEEEcompsocitemizethanks{\IEEEcompsocthanksitem Y. Majdoub is also with the National School of Engineering, Gabes, Tunisia.}%
}

%
%

\markboth{IEEE TRANSACTIONS ON SOFTWARE ENGINEERING}%
{Majdoub \MakeLowercase{\textit{et al.}}: Understanding and Mitigating Library-Related
Issues in LLM-Generated Code}
%



\maketitle

\begin{abstract}
Software practitioners increasingly rely on Large Language Models (LLMs) to generate code that integrates external libraries. However, LLMs often produce incorrect library usage, such as invalid imports, outdated API calls, and hallucinated dependencies, leading to compilation or runtime failures that reduce the reliability of AI-assisted software development.
In this paper, we propose an agentic approach to mitigate library-related errors in LLM-generated code. More specifically, we first conduct an exploratory study to characterize the library-related issues produced by LLMs. Our analysis of 100 LLM-generated code files reveals that 84\% of generated files contain at least one library-related error, with recurring patterns including incorrect import paths, missing imports, hallucinated libraries, deprecated library usage, and unused imports. Based on these findings, we design an agentic approach that integrates task analysis, documentation grounding, code generation, and automated validation to improve library usage during code synthesis.
We evaluate our approach on 300 code generation tasks derived from real-world implementations of rapidly evolving Python frameworks, including LangChain and AutoGen, across five LLMs: GPT-5, DeepSeek-V3, Qwen3, Mistral, and Llama~3. The results show that our approach consistently improves code generation quality across all evaluated models, reducing library-related errors by 38.1\%--54.6\% and increasing code correctness by up to 16\%.
\end{abstract}


%
\IEEEpeerreviewmaketitle

\section{Introduction}
\label{sec:introduction}
 
LLMs have transformed software engineering by enabling new forms of automation across various development activities, including code generation~\cite{joel2025surveyllmbasedcodegeneration}, program repair~\cite{Majdoub2024Debugging}, and testing~\cite{wang2024softwaretestinglargelanguage}.
Among these applications, code generation has received particular attention due to the ability of LLMs to translate natural language descriptions into executable programs~\cite{huynh2025largelanguagemodelscode}. However, generating functionally correct code requires not only producing valid program logic but also correctly interacting with external libraries and their APIs~\cite{abbassi2025taxonomyi}.

A major challenge in this context is that libraries are not static artifacts. They continuously evolve through version updates, refactoring, and module reorganization, causing previously valid usage patterns to become obsolete~\cite{zhuo2025identifyingmitigatingapimisuse}. Since LLMs are pre-trained on historical code repositories, their knowledge may not reflect the current state of rapidly evolving libraries. As a result, generated code may contain incorrect library imports, outdated APIs, or even nonexistent modules~\cite{abbassi2025taxonomyi}. 

Prior work shows that 4.6\% of the libraries appearing in generated code cannot be resolved automatically due to structural mismatches between import names and installable packages~\cite{latendresse2025robustllm}. Although this percentage may appear limited, such issues often result in complete execution failures because a single incorrect dependency can prevent the entire program from running. These failures require developers to manually investigate library versions and compatibility issues, reducing the productivity gains expected from LLM-based development.

This problem is amplified by the rapid evolution of modern software libraries, which frequently introduce API refactoring, module reorganization, and backward-incompatible changes~\cite{zhuo2025identifyingmitigatingapimisuse}. The mismatch between the static code knowledge learned during LLM training and the continuously changing state of libraries often leads to plausible-looking but incorrect code. This issue is particularly severe for fast-evolving frameworks used in LLM-powered applications, such as LangChain~\cite{chase2022langchain} and AutoGen~\cite{wu2023autogenenablingnextgenllm}, where frequent API changes can cause generated code to fail due to outdated imports, missing dependencies, or deprecated APIs. 
Any incorrect library import can turn into total failure, as downstream components rely on precise module structures~\cite{krishna2025importingphantomsmeasuringllm,tambon2024bugslargelanguagemodels}. Unfortunately, leading LLMs such as GPT-5.2 fail in resolving these issues, leaving developers to debug confidently generated but fundamentally broken code~\cite{huang2024largelanguagemodelsselfcorrect}.

To understand and quantify this problem, we first conduct an exploratory study of LLM-generated code files. We collect 100 Python code files generated for real-world programming tasks involving external libraries and manually analyze the encountered library-related failures. From this analysis, we derive five dominant error categories: Incorrect Import Paths, Missing Imports, Hallucinated Imports, Deprecated Imports, and Unused Imports. The results shows that 84\% of generated files contain library-related errors, with an average of 2.7 errors per faulty file. These findings suggests that library-related issues represent a significant barrier to reliable LLM-based code generation and motivate the need for approaches that explicitly address library evolution and API compatibility.

After quantifying these issues, we propose an agentic system designed to mitigate library-related errors in LLM-generated code. Our approach grounds code generation in up-to-date documentation through a closed-loop pipeline composed of four main components: an Analyzer agent that identifies required dependencies, collects and filters relevant API documentation using a Documentation Retrieval module , a Code Generator agent that produces documentation-grounded implementations, and a Validator agent that detects and corrects inconsistencies through feedback from a Compilation and Dependency Management module. This design enables continuous validation of library usage against current APIs and reduces reliance on outdated model knowledge.

We evaluate our system on 300 code generation tasks derived from real-world implementations of fast-evolving Python frameworks and compare it against direct prompting across five LLMs: GPT-5~\cite{singh2025openaigpt5card}, DeepSeek-V3~\cite{deepseekai2025deepseekv3technicalreport}, Qwen3~\cite{yang2025qwen3technicalreport}, Mistral~\cite{jiang2023mistral7b}, and Llama3~\cite{grattafiori2024llama3herdmodels}. The results show that our approach consistently improves code correctness across all evaluated models, reducing library-related errors by up to 54.6\% and increasing correctness rates by up to 16\%. Furthermore, to understand the contribution of each component to the overall effectiveness of the system, we further conduct an ablation study by removing individual modules from the complete pipeline while keeping the remaining components unchanged. The results show that each component contributes to reducing library-related errors and improving code reliability, highlighting the importance of combining dependency analysis, documentation retrieval, and iterative validation.

\textbf{Paper Organization.} The rest of this paper is organized as follows: Section~\ref{sec:motivation} presents a motivating example, Section~\ref{sec:exploratory} details our error taxonomy and exploratory study, Section~\ref{sec:solution} introduces our agentic architecture, Section~\ref{sec:experiment} describes the experimental methodology, Section~\ref{sec:results} presents the evaluation results, Section~\ref{sec:discussion} discusses the implications and limitations of our findings, Section~\ref{sec:threats} outlines possible threats to validity, Section~\ref{sec:related} discusses related work, and Section~\ref{sec:conclusion} concludes with a summary of contributions.

\section{Motivating Example}
\label{sec:motivation}
To illustrate the problem of incorrect library usage and imports in LLM-generated code, in this section, we begin with a motivating example.
We prompt ChatGPT to generate Python code implementing a Retrieval-Augmented Generation (RAG) system that operates on a PDF document using the prompt shown in Listing~\ref{lst:rag_prompt}. 
\begin{lstlisting}[style=promptstyle,
caption={Example code generation task used in the evaluation},
label={lst:rag_prompt}]
Write a Python program that uses LangChain to implement 
a RAG system that takes a PDF file as input and answers 
questions strictly based on its content.
\end{lstlisting}
\vspace{2mm}
RAG pipelines are representative of modern application development: they rely on multiple third-party libraries, require correct composition of modular components, and are commonly used in real-world LLM-based systems. We used GPT-5.2, which describes itself as capable of reasoning, code generation, and retrieval-augmented tasks, and which reports a knowledge cutoff in August 2025.

\begin{figure}[t]
\centering
\includegraphics[width=\linewidth, height=2.6cm]{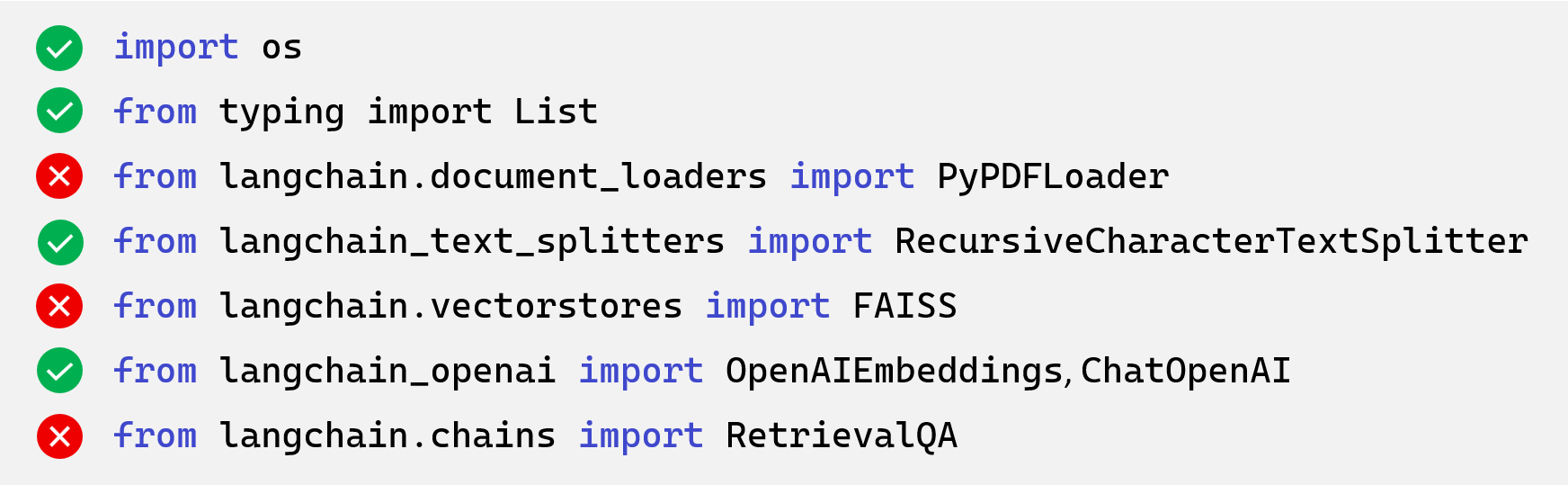}
\caption{Import statements generated by ChatGPT for a LangChain-based RAG system.}
\label{fig:import-error}
\vspace{-5mm}
\end{figure}

In response, ChatGPT generated approximately 60 lines of Python code implementing a standard RAG pipeline. Figure~\ref{fig:import-error} shows the complete set of import statements generated by ChatGPT for this example. Although the generated code correctly identifies the LangChain components typically required to implement such a system, three of the generated import statements are incompatible with the current version of the library. These issues include incorrect module paths and outdated package locations, preventing the generated program from executing successfully.

A closer analysis of these errors reveals that most are caused by incorrect or outdated module paths. For instance, in recent versions of LangChain, the \inlinecode{\small PyPDFLoader} class must be imported from \inlinecode{langchain\_community.document\_loaders}, whereas the generated code attempts to import it from \inlinecode{langchain.document\_loaders}. Similar issues affect other components, including question-answering and vector store utilities. Notably, the imported symbols themselves are valid LangChain abstractions; the errors arise solely from changes in how these components are organized within the library.

This example highlights a broader challenge in automating code generation when programs depend on rapidly evolving libraries. 
Modern frameworks such as LangChain frequently undergo refactoring, modularization, and API reorganization, causing previously valid import paths and usage patterns to become obsolete across versions. 
Although the underlying functionality may remain available, changes in library structure can cause the generated code to fail during execution. 

These failures can be attributed to two main factors. 
First, LLMs may hallucinate libraries, modules, or API usage patterns that do not exist in the target environment. Second, they rely on pre-trained knowledge acquired from historical code and documentation, which may not reflect the current state of rapidly evolving libraries. In both cases, the generated code may contain incorrect dependencies or API usages, leading to execution failures.

Motivated by this example, this work has two main research goals. First, we investigate the types of import-related errors that LLMs produce in generated code, with the objective of identifying and classifying recurring error patterns, particularly in rapidly evolving libraries. Second, we develop a novel approach to fix import errors in the generated code by enabling the model to autonomously identify import failures and recover by discovering and applying the correct import statements using available documentation.

\section{Exploratory Study}
\label{sec:exploratory}

To better understand the limitations of LLM-generated code when interacting with external libraries, we conduct an exploratory study that systematically analyzes generated programs to characterize and quantify library-related import issues, particularly those caused by the use of fast-evolving frameworks. The objective of this study is to identify common patterns of library misuse and measure their prevalence across generated files. Understanding these issues is essential for designing approaches that can automatically detect and mitigate library-related failures. Furthermore, quantifying the frequency of different error patterns allows us to identify the most dominant failure modes and provides insights into the challenges that should be addressed to improve the reliability of LLM-based code generation.

\subsection{Dataset Creation}
\label{sec:exploratory_dataset}
To the best of our knowledge, there is no publicly available dataset that captures LLM code generation behavior using real-world reference implementations. Existing benchmarks often rely on fixed-version libraries, which may not capture the challenges introduced by evolving APIs and abstractions. To address this gap, we curated a dataset of 100 prompt--code pairs through a three-stage process: repository selection, prompt derivation, and LLM-based generation.

First, we collected open-source GitHub repositories using LLM-centric frameworks, such as LangChain, AutoGen, and CrewAI, to derive realistic programming tasks. Using the GitHub REST API, we filtered the retrieved repositories to retain public Python projects containing executable source code while excluding forks and repositories that did not satisfy our predefined selection criteria. This process yielded approximately 400 repositories. From this collection, we manually selected 100 repositories that contained concrete usage of the target frameworks and relevant API interactions. These repositories were used to identify realistic scenarios, API usage patterns, and task requirements for prompt construction.

Following repository selection, we transformed functionality from the reference implementations into natural language prompts. Each prompt was designed to clearly specify the task requirements, expected behavior, and relevant constraints~\cite{sasaki2025landscape}. We use ChatGPT for this step due to its demonstrated effectiveness in software engineering tasks and its ability to translate code functionality into natural language descriptions~\cite{sridhara2023chatgptstudyutilityubiquitous}. Each resulting prompt was then manually reviewed by two of the authors and refined to ensure clarity, completeness, and consistency with the reference implementation. This validation step ensures that the prompts are understandable to a human developer and accurately reflect the functionality implemented in the corresponding reference code. An example of a resulting prompt is shown in Listing~\ref{lst:dataset_prompt}.

\begin{lstlisting}[style=promptstyle, caption={Example of a generated dataset prompt}, label={lst:dataset_prompt}]
Create a video transcription and translation system using 
Whisper AI and OpenAI. Implement recognize_transcript_from_
video function that uses Whisper small model to transcribe 
audio with timestamps, saves to "transcription.txt". Add 
translate_text function using OpenAI API for subtitle 
translation. Include translate_transcript function that 
processes the transcription file and creates translated 
versions. Use proper error handling, dotenv for environment 
variables, and moviepy imports.


\end{lstlisting}

In the final stage, the curated prompts were provided to LLMs, including Mistral and DeepSeek, to generate corresponding code solutions. These outputs form the basis of our evaluation, while the original repository implementations serve as reference implementations for assessing the generated solutions. The resulting dataset of 100 prompt--code pairs covers application scenarios such as data ingestion pipelines, multi-agent coordination workflows, tool calling, retrieval-augmented generation, and external API integration.

\subsection{Annotation Process}
To systematically analyze library-related errors in LLM-generated code, we developed a structured annotation process in which we identify, categorize, and quantify the 
types of issues occurring in import statements and library usage.

\subsubsection{Preliminary Thematic Analysis} Prior to formal annotation, the two annotators jointly reviewed a randomly selected subset of 25 code samples and discussed each sample to identify recurring patterns and types of library-related errors. This preliminary thematic analysis was conducted to establish a common understanding of the error types present in LLM-generated code, particularly given that agentic frameworks frequently undergo breaking changes to their APIs and module structures. Through discussion and comparison, the annotators identified recurring error patterns, refined guideline interpretations, and established a shared consensus for ambiguous cases (e.g., valid imports using deprecated APIs). These finalized themes and criteria guided all subsequent annotations.

\begin{table*}[t]
\centering
\caption{Taxonomy and distribution of import-related errors observed in LLM-generated code.}
\label{tab:import-taxonomy}
\small
\renewcommand{\arraystretch}{1.3}%
\begin{tabular}{p{1.5cm} p{5.8cm} p{5.5cm} c c}
\toprule
Issue & Description & Example & Instances & Files Affected \\
\midrule
Incorrect     Import Path &
The imported library, class, or function exists, but the specified module or submodule path does not match the actual library structure. &
Using \inlinecode{from langchain.chat\_models import } \inlinecode{ChatOpenAI} where 
\inlinecode{ChatOpenAI} is defined in \inlinecode{langchain\_openai}. &
93 & 81 \\
\midrule
Missing Import &
A required import statement is absent despite the component being referenced elsewhere in the code. &
AutoGen components are referenced without importing the package using \inlinecode{import autogen}. &
79 & 54 \\
\midrule
Hallucinated Import &
Imports of non-existent libraries, modules, or classes that appear syntactically valid but do not correspond to any real API elements. &
\inlinecode{from openai import OpenAIGym} OpenAIGym does not exist in the OpenAI library. &
28 & 25 \\
\midrule
Deprecated Import &
Reliance on outdated or superseded APIs that are incompatible with current libraries versions. &
Using \inlinecode{import openai} instead of the updated client-based API
\inlinecode{from openai import OpenAI}. &
17 & 17 \\
\midrule
Unused Import &
An import statement that is syntactically valid but never referenced in the generated code, introducing unnecessary dependencies and reducing code clarity. &
\inlinecode{import requests} is included in the script, but no functions or classes from \inlinecode{requests} are used. &11&9\\
\bottomrule
\end{tabular}
\end{table*} 

\subsubsection{Independent Review of Generated Code} To maintain high annotation quality and consistency, the remaining dataset was divided into three equal batches ($n = 25, 25, 25$). Two annotators independently evaluated each file by cross-referencing the generated code against the original prompt and official documentation. Each code sample was evaluated in the context of its associated prompt, examining the referenced packages, modules, classes, and functions and their integration into the generated implementation. Evaluation criteria included:

\begin{itemize}
    \item \textit{Package Existence:} Each top-level package was verified using documentation to ensure that it is published and correctly spelled. Third-party libraries were also cross-referenced with official documentation or reputable online sources \cite{krishna2025importingphantomsmeasuringllm}.

    \item \textit{Path Structure and Class Verification:} Imported submodules and classes were checked against official library documentation or GitHub repositories to confirm their existence and correct hierarchy. For instance, an import such as \inlinecode{from langchain\_core. documents import Document} was validated by confirming that \inlinecode{langchain\_core.documents} is a valid module and that \inlinecode{Document} is defined within it \cite{chen2025mitigatingapihallucinationcode}.

    \item \textit{Cross-Reference with Reference Code:} Imports were compared against reference implementations to identify structural differences, missing components, or case-sensitivity errors \cite{austin2021programsynthesislargelanguage}.

    \item \textit{Import Usage:} The code was examined to determine whether imported libraries and modules were actually referenced in the generated code, allowing unused and missing imports to be identified \cite{jesse2023large}.
\end{itemize}

\subsubsection{Taxonomy Construction and Classification} Following the initial exploratory review, recurring error patterns were synthesized into a formal taxonomy of import-related error categories. Using an open thematic analysis approach, annotators reviewed code samples to identify underlying error mechanisms, which were iteratively compared, grouped, and refined into a standardized schema. The resulting taxonomy abstracted specific failure modes into high-level analytical categories (e.g., structural path errors versus semantic hallucinations) and provided a consistent framework for subsequent annotation and quantitative analysis.

\subsubsection{Inter-Annotator Agreement Assessment} During batch annotation, annotators independently classified code samples using the established taxonomy. After each batch, they held review meetings to resolve disagreements by consensus and refine annotation guidelines to improve consistency. Annotation quality was assessed by comparing results for each file across three agreement levels: Full Agreement, where both annotators identified the same issues, classifications, and error counts; Partial Agreement, where annotators agreed on most findings but differed on one or two classifications or identifications; and No Agreement, where findings differed substantially across multiple classifications, identifications, or error counts.

\subsection{Study Results}
\label{sec:expl_results}

We present the results of our exploratory study based on the 100 annotated code files. To assess the reliability of our classification, we measured inter-annotator agreement using Cohen’s kappa \cite{mchugh2012interrater}. The annotators achieved a Cohen’s kappa of $\kappa = 0.86$, indicating a very high level of agreement. For the small number of cases with initial disagreements, a joint reconciliation session was conducted to reach final consensus. We analyze the distribution of errors across our dataset of 100 evaluable prompt--code pairs, excluding 7 cases where no code was generated. Reported frequencies refer to individual error instances rather than unique files unless otherwise noted.

A critical finding of this study is the high density of errors within individual files:
\begin{itemize}
    \item \textit{High Failure Prevalence:} Of the 100 code files analyzed, 84 files contained at least one import-related error.
    \item \textit{Error Density \& Clustering:} Within the 84 problematic files, we recorded a total of 228 distinct error instances, averaging approximately 2.7 errors per failing file.
\end{itemize}

We introduce a taxonomy of import-related errors observed in LLM-generated code for rapidly evolving libraries. The taxonomy was developed during preliminary exploration and finalized prior to large-scale annotation, serving as the shared framework for all subsequent analysis. It consists of five categories (Table~\ref{tab:import-taxonomy}), each capturing a distinct deviation from correct, complete, and prompt-aligned import statements. The categories are not mutually exclusive, and a single code file may exhibit multiple error types.

These errors span both structural and semantic dimensions of library usage: \textit{incorrect import paths} and \textit{missing imports} reflect mismatches with actual module organization, while \textit{hallucinated imports} and \textit{deprecated imports} indicate discrepancies between the model’s internal knowledge and the current documented API. Such issues are particularly prevalent in agentic frameworks such as LangChain, AutoGen, and CrewAI, which differ from stable software ecosystems due to their rapid evolution and frequent breaking changes to internal hierarchies and core abstractions \cite{twist2025libraryhallucinationsllmsrisk, krishna2025importingphantomsmeasuringllm}. 

Across the 228 recorded error instances, failures were distributed across five key categories (Table~\ref{tab:import-taxonomy}):

\begin{enumerate}
\item \textit{Incorrect import paths (40.8\%):} The most prevalent error type. In these cases, the model frequently selected the correct high-level library but failed to correctly navigate its internal module hierarchy, such as placing a class or function in an incorrect submodule. For example, the model generated \texttt{from langchain.chat\_models import ChatOpenAI}, although \texttt{ChatOpenAI} is defined in \texttt{langchain\_openai}. This pattern reflects the inherent instability of agentic framework architectures, where internal paths are frequently reorganized across versions. These libraries often undergo frequent API revisions and structural refactorings, similar to broader patterns seen in evolving software ecosystems where dependency networks grow in complexity and interface changes challenge developers \cite{10.1145/3470133}. The prevalence of this error type suggests that LLMs struggle not with recognizing which libraries to use, but with maintaining accurate, fine-grained knowledge of their internal structure. 

\item \textit{Missing imports (34.6\%):} Represented the second largest category. These errors occurred when required dependencies or framework components were omitted despite being necessary for execution. For example, AutoGen components were referenced in the generated code without including the required \texttt{import autogen} statement. Notably, a portion of these cases involved missing core components from the target frameworks (e.g., LangChain, AutoGen, CrewAI), highlighting an increased challenge when dealing with recently updated or less established ecosystems that may have evolving APIs. These findings suggest that LLMs may struggle to track the complete set of dependencies required for correct execution, particularly in frameworks with complex and rapidly changing abstractions.

\item \textit{Hallucinated imports (12.3\%):} The third most frequent error type reveals a tendency for the model to fabricate plausible-sounding modules or classes when faced with uncertainty. For example, the model generated \texttt{from openai import OpenAIGym}, although \texttt{OpenAIGym} does not exist in the OpenAI library. Unlike incorrect paths, which often require minor corrections, hallucinated imports represent a more severe failure mode, as they reference non-existent components that cannot be resolved without external documentation or trial-and-error debugging. This highlights the limitations of relying solely on static model knowledge when generating code for fast-evolving agentic ecosystems.

\item \textit{Deprecated imports (7.5\%):} Illustrated the lag between the model's internalized knowledge and the current state of rapidly changing APIs. For example, the model generated \texttt{import openai} instead of the updated client-based API usage \texttt{from openai import OpenAI}. In these cases, the imports were often syntactically valid in older versions of the library but contradicted the prompt’s intent to use newer, client-based or modularized interfaces. 

\item \textit{Unused imports (4.8\%):} Observed reflecting inefficiencies that, while less severe, still contribute to reduced code clarity and maintainability. For example, the model included \texttt{import requests} in the generated script, but no functions or classes from \texttt{requests} were used.
\end{enumerate}

The observed error distributions reflect the architectural complexity and rapid evolution of agentic frameworks, which differ from more stable software ecosystems \cite{derouiche2025agenticaiframeworksarchitectures}. In the absence of stable usage patterns, models are more likely to fabricate plausible but unsupported imports or rely on outdated interfaces. Collectively, these findings show that import-related errors in agentic framework code are systematic, reflecting limitations in generating version-consistent code for rapidly evolving libraries and highlighting the need for library-aware approaches that can improve models’ ability to leverage accurate, up-to-date library information during code generation.

\section{Approach for Solving Library-Related Issues}
\label{sec:solution}

\begin{figure}[t]
\vspace{8mm}
\centering
\includegraphics[width=\linewidth, height=4cm]{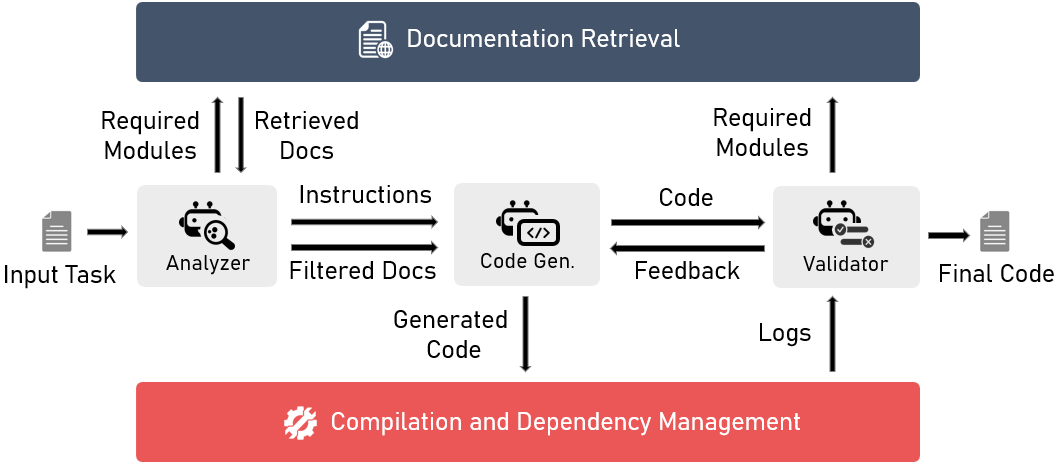}
\caption{Overview of the proposed approach}
\vspace{2mm}
\label{fig:system}

\end{figure}

To address the problem of incorrect external library usage in LLM-generated code, we propose a novel agentic system for library-aware code generation that formulates code generation as an iterative and structured process. An overview of the system is presented in Figure~\ref{fig:system}. 
The system takes as input a natural language task describing the desired functionality of the program. It then begins with task analysis and libraries documentation retrieval, proceeds through code generation, and finishes with automated validation and correction. The final output is a validated code solution that satisfies the task requirements while solving library-related issues.

This workflow is implemented via three interacting agents: \textit{Analyzer} agent, \textit{Code Generator} agent, and \textit{Validator} agent, each responsible for a specific stage of the process while communicating through shared modules for \textit{Documentation Retrieval} and \textit{Compilation}. We present more details about these agents in the following subsections.

\subsection{Analyzer}

The \textit{Analyzer} prepares the information required for code generation. It receives as input a natural language code generation prompt describing the functionality of the program to be implemented, such as the prompt shown in Listing~\ref{lst:rag_prompt}. Based on this input, it first identifies the external libraries and modules required to implement the requested functionality, then retrieves and filters their corresponding documentation. The \textit{Analyzer} performs these operations in two stages.

\subsubsection{Dependency Analysis.} The first task consists of identifying the external libraries required to implement the requested functionality. To achieve this, the \textit{Analyzer} is prompted to analyze the semantic requirements of the input task and identifies the required third-party libraries, frameworks, and submodules while explicitly excluding Python standard library modules.
The prompt instructs the agent to return the extracted dependencies according to a predefined JSON schema, ensuring that the output can be reliably processed by subsequent components. The generated JSON representation contains the identified libraries, their associated modules, and their intended usage within the generated program, as illustrated in Listing~\ref{lst:dependency_schema}.

\begin{lstlisting}[style=promptstyle, caption={Output format of dependency specification stage}, label={lst:dependency_schema}]
{"dependencies": [{
   "library": "library_name",
   "modules": [ "module1", "module2"],
   "purpose": "how the library is used"
}]}
\end{lstlisting}
\vspace{2mm}

\subsubsection{Documentation Processing.} Using the generated dependency specification, \textit{Analyzer} interacts with the \textit{Documentation Retrieval} module to retrieve the official documentation corresponding to the identified libraries. Since the retrieved documentation may contain redundant or task-irrelevant information, \textit{Analyzer} performs a filtering step to retain only the documentation relevant to the identified dependencies. Specifically, it removes duplicated and unrelated content while preserving verified import statements, API descriptions, and usage examples associated with the required modules. The prompt instructs the LLM to return the extracted dependencies according to a predefined JSON schema, ensuring that the output can be reliably processed by subsequent components. The generated JSON representation contains the identified libraries, their associated modules, and their intended usage within the generated program, as illustrated in Listing~\ref{lst:dependency_schema}.

\begin{lstlisting}[style=promptstyle, caption={Output format of documentation processing stage}, label={lst:dependency_schema}]
{"libraries": [{
    "name": "library_name",
    "imports": ["from module import ClassName"],
    "usage_examples": ["example code snippet"],
    "relevant_information": [ "usage details"]
}]}
\end{lstlisting}

\subsection{Documentation Retrieval}
\label{sec:documentation_retrieval}

The \textit{Documentation Retrieval} module supports \textit{Analyzer} by collecting authoritative and task-relevant API documentation for the dependencies identified during task analysis. Given the structured dependency specification produced by \textit{Analyzer}, the module queries trusted web sources using the \inlinecode{DuckDuckGo} Search API and retrieves official documentation pages corresponding to the required libraries and submodules. 
Rather than exposing raw documentation to subsequent stages, the \textit{Documentation Retrieval} module processes the retrieved pages to extract concrete artifacts that are directly useful for code generation. 

First, the module parses the HTML content using \inlinecode{BeautifulSoup} and identifies code blocks and text segments containing API usage information. It then extracts import statements and usage patterns from these extracted sections by analyzing the code examples provided in the documentation. For instance, it identifies valid imports such as \inlinecode{from langchain.tools import tool} and retains short code snippets that demonstrate the correct usage of the corresponding APIs.
These patterns capture common operations relevant to agent-based and LLM-driven workflows, such as initializing chains, configuring retrievers, instantiating agents, and invoking LLM calls, while discarding explanatory text and unrelated examples. The output of this module is an intermediate documentation bundle consisting of two components: (i) a list of verified import statements and (ii) a collection of concise, executable usage snippets.

Once retrieval is complete, the documentation bundle is returned to \textit{Analyzer}, which applies additional task-specific filtering and alignment. This final refinement step ensures that only documentation consistent with the inferred task intent is retained, eliminating outdated, redundant, or incompatible API references before the bundle is passed to \textit{Code Generator}.

\subsection{Code Generator}

After receiving the structured generation instructions and the filtered library documentation produced by \textit{Analyzer}, \textit{Code Generator} synthesizes a complete executable implementation. The agent is explicitly constrained in the prompt to follow the provided import statements, module paths, and usage patterns extracted from documentation results. This constraint enforces alignment with official APIs and mitigates common failure modes such as hallucinated, deprecated, or incompatible library usage.

The resulting output is a full candidate implementation that is not yet guaranteed to be correct or executable in practice. The system therefore treats this output as a provisional solution and forwards it to the validation stage, where the generated code is verified for syntactic validity and executability through the \textit{Compilation} module and subsequently examined by \textit{Validator} to assess semantic correctness, API conformance, and consistency with the original task specification.

\subsection{Compilation and Dependency Management}
\label{sec:environment_preparation}

The \textit{Compilation and Dependency Management} module assists \textit{Validator} by performing systematic checks that ensure the generated code is syntactically valid and supported by the execution environment. Upon receiving a candidate implementation from \textit{Code Generator}, the module compiles the generated code without executing its program logic. If compilation fails, the resulting error logs are provided to \textit{Validator}, which analyzes the errors together with the task requirements and retrieved documentation to determine whether the generated solution requires correction and to formulate appropriate feedback for the next generation iteration.

When the compilation succeeds, the module extracts all import statements from the source code and verifies whether the corresponding package is available in the execution environment. Missing dependencies are automatically installed using \inlinecode{pip} through the \inlinecode{subprocess} module after verifying their availability on PyPI. For instance, if \inlinecode{langchain} is missing, the system attempts to install it by running a \inlinecode{pip install langchain} command in a subprocess and then verifies that the module can be successfully imported using \inlinecode{importlib}.

Once all required dependencies are confirmed to be available, the module compiles the code again to ensure that the code remains valid under the fully prepared environment. Throughout this process, detailed logs are collected, summarizing previously installed packages, newly installed packages, unresolved dependencies, and any compilation errors encountered. These logs are then forwarded to \textit{Validator} to support higher-level validation and decision-making.

\subsection{Validator}

The Validator agent serves as the quality assurance component of our agentic pipeline, responsible for systematically evaluating the functional correctness and library conformance of the code produced by the Code Generator. Upon receiving a candidate implementation, the Validator first inspects the compilation logs and dependency resolution status provided by the Compilation module. This includes verifying whether all required third-party packages were successfully installed, whether import statements resolved to actual modules in the current environment, and whether any runtime errors—such as \inlinecode{ModuleNotFoundError} or \inlinecode{AttributeError}, were triggered during attempted execution. 

Next, the Validator cross-references the generated imports and API calls against the filtered documentation bundle prepared earlier by the Analyzer. For instance, if the code uses \inlinecode{from langchain.chat\_models import ChatOpenAI}, the Validator checks whether this path exists in the current LangChain documentation. If it has been relocated (e.g., to \inlinecode{langchain\_openai.ChatOpenAI}), the Validator flags this as an error. In addition to structural validation, the Validator assesses semantic alignment with the original task specification. It verifies that the chosen libraries and components fulfill the functional requirements implied by the natural language prompt. This prevents situations where the code compiles successfully, but fails to implement the intended behavior due to incomplete or misconfigured library usage. 
When issues are detected, the Validator constructs structured, actionable feedback. This feedback includes issue category, the exact location in the code, a description of the mismatch, and specific correction instructions. The feedback is then passed back to the Code Generator, which uses it to regenerate only the affected portions of the code while preserving the rest of the implementation. 

The entire loop repeats iteratively until the solution satisfies the validation criteria or reaches the configured maximum number of iterations. This iteration limit can be adjusted depending on the desired trade-off between refinement capability and computational cost, preventing infinite refinement cycles.
At each iteration, the Validator re-evaluates the updated code using the same rigorous criteria, ensuring progressive convergence toward a fully correct and executable solution.

\section{Experimental Setup}
\label{sec:experiment}

To evaluate our approach, we address the following research questions (RQs):

\begin{itemize}

\item[\small RQ1] \textit{How effective is the proposed system in solving library-related issues across different LLMs?}

To measure the effectiveness of our approach, we evaluate it on a dataset of 300 code generation tasks. To put the results into context, we compare it against a baseline where we use direct prompting. We assess both approaches using the code correctness rate, the total number of library-related errors, and the error reduction rate across five language models. We also analyze the reduction achieved for individual library-related error categories.

\item[\small RQ2] \textit{How do the individual components of the proposed system contribute to its overall effectiveness?}

To understand the contribution of each component to the system performance, we perform an ablation study in which we compare variants of the system by removing one component from the complete design while keeping all other components unchanged. Specifically, we evaluate the impact of removing the Analyzer, Documentation Retrieval, Compilation and Dependency Management, and Validator modules under the same experimental conditions, measuring their individual contributions to reducing library-related errors and improving code executability.

\end{itemize}

In the following, we present the dataset construction process, evaluation metrics, and experimental configurations used in the evaluation.

\subsection{Evaluation Dataset}

To the best of our knowledge, no existing dataset is specifically designed to evaluate the ability of LLM-based systems to detect and resolve library-related errors caused by rapidly evolving APIs. Therefore, we curate a new dataset to evaluate our proposed approach.
We selected five widely adopted and rapidly evolving frameworks used in LLM-powered applications, namely \textit{LangChain}, \textit{AutoGen}, \textit{CrewAI}, \textit{LlamaIndex}, and \textit{Agno}. These frameworks were selected based on their popularity in real-world LLM application development and their frequent evolution, as reported in previous studies~\cite{liu2026largescalestudydevelopmentissues}.

Following the repository selection and prompt-generation methodology described in Section~\ref{sec:exploratory_dataset}, we collected Python implementations from open-source GitHub repositories associated with the selected frameworks. However, to reflect recent library versions and usage patterns, we specifically selected repositories that were created or actively maintained after June 2025. We further retained implementations representing realistic LLM-powered application scenarios whose functionality depends on the correct use of external libraries and APIs.

For each selected implementation, we used ChatGPT to generate a corresponding natural language code-generation task by extracting its intended functionality. As in the exploratory dataset, the prompts describe the expected behavior while intentionally omitting framework-specific class names, import paths, and low-level implementation details. The original implementations were used only as references for task construction and were not provided to the evaluated models.

The prompts were manually reviewed by two annotators to ensure that they accurately represented the intended functionality without exposing implementation-specific information. This design requires the evaluated LLMs to independently determine the appropriate library APIs and implementation rather than reproduce the original source code.

The final dataset contains 300 code-generation tasks covering diverse LLM application scenarios, including agent coordination, tool usage, document processing, and modular pipeline composition.

\subsection{Evaluation Metrics}

We evaluate the generated files using metrics that measure overall correctness and the effectiveness of the proposed system in improving code generation reliability, particularly its ability to mitigate library-related issues.

\begin{enumerate}

\item \textit{Code correctness rate:}
This metric measures the percentage of generated files that successfully compile and satisfy library usage requirements. It is computed as:

\[
\text{Code Correctness Rate}(\%) =
\frac{N_{\text{correct}}}{N_{\text{total}}} \times 100
\]

where $N_{\text{correct}}$ is the number of generated files that has been successfully compiled, and $N_{\text{total}}$ is the total number of evaluated files.

\item \textit{Error reduction rate:}
This metric measures the relative reduction in the number of library-related errors achieved by the proposed system compared with direct LLM prompting. It is computed as:

\[
\text{Error Reduction Rate}(\%) =
\frac{N_{\text{baseline}} - N_{\text{system}}}
{N_{\text{baseline}}} \times 100
\]

where $N_{\text{baseline}}$ and $N_{\text{system}}$ represent the number of library-related errors detected in files generated using direct LLM prompting and the proposed system, respectively.

\item \textit{Median refinement iterations:}
This metric measures the median number of correction cycles required by the proposed system to generate the final solution. 

\end{enumerate}

In addition to these metrics, we perform an error category analysis to examine how the proposed system affects different types of library-related issues.

\subsection{Evaluation Settings}



We evaluate our proposed system using five LLMs: GPT-5, DeepSeek-V3, Qwen3, Mistral, and Llama~3. These models were selected to cover diverse LLM families, including both proprietary and open-weight models, and to represent different levels of code generation capabilities. 

For each model, we compare our approach against direct prompting baseline. In direct prompting, the LLM receives a natural language task description and generates the corresponding code directly. It provides an effective reference point for measuring the additional benefits introduced by integrating external knowledge and automated validation mechanisms. Both evaluation settings use the same dataset, task prompts, and model configurations to ensure a fair comparison.

\section{Results}
\label{sec:results}

This section presents the results of our evaluation according to the defined research questions. We first investigate the effectiveness of the proposed approach in reducing library-related errors and improving the correctness of generated code (RQ1). We then analyze the contribution of individual system components through an ablation study (RQ2).

\begin{table*}[t]
\centering
\caption{Comparison between direct prompting and the proposed system across five LLMs on 300 code generation tasks.}
\label{tab:main_results}
\resizebox{\textwidth}{!}{
\renewcommand{\arraystretch}{1.5}
\begin{tabular}{lcccccccccc}
\toprule
& \multicolumn{5}{c}{Direct Prompting}
& \multicolumn{5}{c}{Our Approach} \\
\cmidrule(lr){2-6}
\cmidrule(lr){7-11}

Metric
& GPT-5
& DeepSeek-V3
& Qwen-3
& Mistral
& Llama-3
& GPT-5
& DeepSeek-V3
& Qwen-3
& Mistral
& Llama 3 \\
\midrule

Overall Correctness
& 81\% & 71\% & 67\% & 64\% & 61\%
& \textbf{85\%} (+5\%) & 83\% (+12\%) & 82\% (+15\%) & 78\% (+14\%) & 77\% (\textbf{+16\%}) \\

\midrule

Median Iterations
& -- & -- & -- & -- & --
& 1.68 & 2.07 & 1.81 & 2.37 & 2.53 \\
\midrule

Total Library Errors
& 168 & 196 & 212 & 302 & 339
& \textbf{104} (-38\%) & 114 (-42\%) & 122 (-42\%) & 137 (\textbf{-55\%}) & 176 (-48\%) \\
\midrule

Missing Import
& 29 & 36 & 31 & 38 & 53
& 22 & 26 & \textbf{19} & 28 & 37 \\

Incorrect Import Path
& 35 & 41 & 54 & 89 & 103
& 31 & \textbf{30} & 36 & 38 & 47 \\

Hallucinated Library
& 28 & 31 & 37 & 20 & 48
& \textbf{11} & 13 & 18 & 13 & 21 \\

Unused Import
& 27 & 19 & 28 & 53 & 44
& 9 & \textbf{7} & 14 & 18 & 23 \\

Deprecated Library 
& 49 & 69 & 62 & 102 & 91
& \textbf{31} & 38 & 35 & 40 & 48 \\

\bottomrule
\end{tabular}
}
\end{table*}

\subsection{Effectiveness of the Proposed Approach}
\label{sec:results1}
\subsubsection{Impact on library-related errors}

Table~\ref{tab:main_results} summarizes the results of our system compared to direct prompting across all models. 
Our proposed system consistently improves overall correctness across all evaluated models, demonstrating the effectiveness of its library-aware retrieval and validation mechanisms. It increases the correctness rate by 4\% for GPT-5, 12\% for DeepSeek-V3, 15\% for Qwen3, 14\% for Mistral, and 16\% for Llama~3. The larger improvements observed for open-weight models indicate that they benefit more from access to external library knowledge and automated validation. However, the improvements achieved by GPT-5 also demonstrate that the proposed approach provides complementary benefits beyond model capability alone.

A similar pattern is observed for overall library-related errors. Across all evaluated models, our system substantially reduces the total number of library-related errors, achieving reductions of 38.1\% for GPT-5, 41.8\% for DeepSeek-V3, 42.5\% for Qwen3, 54.6\% for Mistral, and 48.1\% for Llama~3. 
These results confirm that integrating library-aware retrieval and validation into the generation process substantially enhances code correctness.

The magnitude of the improvements varies across models depending on their initial performance under direct prompting. Models with a higher number of library-related errors, particularly open-weight models, benefit more substantially from the proposed system, achieving reductions exceeding 40\% in total library errors and improvements of up to 16\% in code correctness. This can be attributed to the ability of documentation grounding and validation mechanisms to provide reliable external library knowledge during generation. Nevertheless, GPT-5, which already achieves strong results under direct prompting, still benefits from the proposed system, demonstrating that the approach improves code correctness even for strong proprietary models. These findings highlight that the proposed system enhances LLM-based code generation by integrating up-to-date library knowledge and automated validation rather than merely compensating for limitations in model knowledge.

\subsubsection{Iterative Refinement}

Our system relies on an iterative refinement process, where the generated code is progressively improved based on feedback from validation and compilation stages. To analyze this behavior, we report the median number of refinement iterations required to obtain the final solution for each evaluated model (Table~\ref{tab:main_results}). The results show that the system converges within a limited number of iterations across all models, with median values ranging from 1.68 iterations for GPT-5 to 2.53 iterations for Llama~3.

The variation in refinement steps reflects differences in the initial quality of generated code. Models with stronger code generation capabilities, such as GPT-5, typically require fewer correction cycles, while models that generate more library-related issues require additional iterations to incorporate validation feedback and correct detected inconsistencies. Overall, the limited number of required iterations demonstrates that the proposed feedback loop improves code reliability without introducing excessive refinement overhead.

\begin{figure}[t]
\centering
\includegraphics[width=\linewidth, height=4cm]{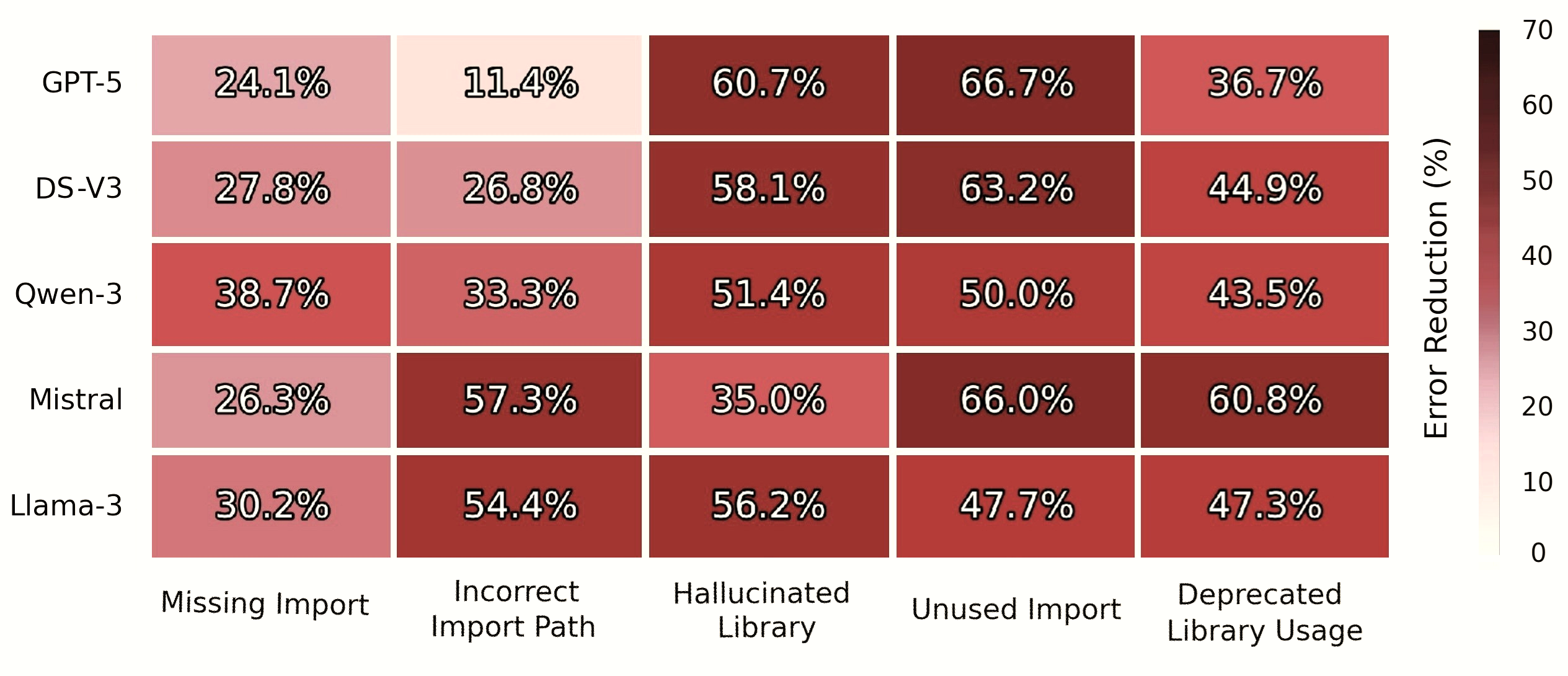}
\caption{Library-related error reduction achieved by the proposed system across error categories and evaluated models.}
\label{fig:heatmap}
\end{figure}

\subsubsection{Effects across library-related error categories}

Table~\ref{tab:main_results} presents a detailed comparison of the proposed system and direct prompting across different categories of library-related errors. The results show that the proposed system consistently reduces all error categories across the evaluated models. The largest improvements are observed for hallucinated libraries, deprecated library usage, and incorrect import paths, which represent challenging failures caused by outdated or incomplete library knowledge.
For hallucinated libraries, the proposed system substantially reduces errors across all models. For example, the number of hallucinated library errors decreases from 31 to 13 for GPT-5 (60.7\% reduction) and from 48 to 21 for DeepSeek-V3 (58.1\% reduction). Similar reductions are observed for the remaining models, demonstrating that grounding code generation with retrieved documentation helps prevent the generation of non-existent libraries.

The proposed system also significantly mitigates deprecated library usage. For instance, Mistral shows a reduction from 102 to 40 errors (60.8\%), while Llama~3 decreases from 91 to 48 errors (47.3\%). These improvements indicate that access to updated documentation enables the system to replace outdated API patterns with currently supported alternatives. Moreover, incorrect import paths, unused imports, and missing imports are consistently reduced across models, confirming the effectiveness of the proposed approach in improving dependency resolution and library integration.

Figure~\ref{fig:heatmap} further provides a complementary visualization of the reduction rates achieved across error categories and evaluated models. The heatmap highlights that the improvements are not limited to a specific model or error type. Instead, the proposed system consistently reduces multiple categories of library-related failures, with particularly strong reductions observed for hallucinated libraries and deprecated usage patterns. These results demonstrate that combining documentation retrieval with validation mechanisms provides broad protection against diverse library-related issues rather than addressing only individual import failures.

Overall, the results confirm that our system provides broad improvements across different LLMs and error categories rather than being limited to a specific model architecture.
These findings demonstrate that the system does not only correct isolated import statements, but systematically improves the use of external libraries throughout the generation process. By combining documentation retrieval with post-generation validation, the system addresses multiple failure modes simultaneously, including incorrect imports, hallucinated libraries, and deprecated APIs.
\begin{tcolorbox}[arc=0pt,colback=gray!5,colframe=black!70,title=Answer to RQ1,top=2mm,bottom=2mm,right=2mm,left=2mm]
Our system consistently improves LLM-generated code quality across all evaluated models, with GPT-5 achieving the best overall performance. Compared with direct prompting, our system increases correctness by up to \textbf{16\%}, reduces library-related errors by up to \textbf{54.6\%}, and mitigates major error categories, including incorrect imports, hallucinated libraries, and deprecated API usage. Furthermore, the iterative refinement process resolves detected issues within a limited number of correction cycles, showing an effective balance between improvement and refinement cost.
\end{tcolorbox}

\subsection{Contribution of System Components}
\label{sec:results2}
To investigate the contribution of each component to the overall effectiveness of the proposed system, we conduct an ablation study using DeepSeek-V3. DeepSeek-V3 was selected for this study as a representative open-weights model because it achieved the strongest overall performance and largest error reduction among open models in RQ1, making it an ideal testbed to observe how modular components impact system capability. Specifically, we compare the complete system against four variants, each obtained by removing one optional component while keeping the remaining components unchanged. Since the Code Generation module represents the core mechanism responsible for producing the final program, it remains enabled in all configurations. This analysis evaluates the individual contribution of the Analyzer agent, Validator agent, Documentation Retrieval module, and Compilation module in improving code correctness and reducing library-related errors. The results are presented in Table~\ref{tab:ablation_results}.

The results demonstrate that each component contributes to the overall performance of the proposed system. Removing any individual component leads to a decrease in correctness rate and an increase in library-related errors compared with the complete configuration. The full system achieves the best performance, reaching a correctness rate of 83\%, while producing the lowest number of library-related errors (114) and affected files (51). These results confirm that the effectiveness of the proposed approach relies on the complementary interaction between its different components.

The Documentation Retrieval module has the largest impact on library usage correctness. Removing this component results in the lowest correctness rate among all configurations (75\%) and increases library-related errors from 114 to 177, with 76 affected files. This significant degradation highlights the importance of providing the generation process with up-to-date external library knowledge. In particular, retrieved documentation helps mitigate incorrect imports, deprecated API usage, and hallucinated dependencies that may not be reliably resolved using the model's internal knowledge alone.

The Validator agent also provides a substantial contribution to the final performance. When the Validator is removed, correctness decreases from 83\% to 78\%, while library-related errors increase to 149 and affected files to 65. This demonstrates the importance of post-generation verification in identifying inconsistencies between generated code and library requirements, enabling further correction before producing the final output.

The Analyzer component contributes by improving task understanding and guiding the generation process. Without the Analyzer, correctness decreases to 76\%, with 164 library-related errors and 71 affected files. This degradation indicates that analyzing the task requirements and identifying relevant library usage patterns before generation helps reduce inappropriate API selection and library misuse.

\begin{table}[t]
\centering
\caption{Ablation analysis of the proposed system components using DeepSeek-V3. Each variant removes one component from the complete system while keeping all remaining components enabled.}
\label{tab:ablation_results}
\renewcommand{\arraystretch}{1.3}
\resizebox{\linewidth}{!}{
\begin{tabular}{lcc}
\toprule
Configuration
& Correctness Rate
& Library Error Count
 \\
\midrule

W/O Analyzer
& 76\%
& 164
 \\

W/O Validator
& 78\%
& 149
 \\

W/O Doc. Retrieval
& 75\%
& \textbf{177}
 \\

W/O Compilation
& 81\%
& 132
 \\

\midrule

Full System
& \textbf{83\%}
& \textbf{114}
 \\

\bottomrule
\end{tabular}
}
\end{table}

Finally, disabling the Compilation module results in a correctness rate of 81\%, with 132 library-related errors and 56 affected files. The limited decrease in correctness (2\%) suggests that current LLMs are generally capable of generating syntactically valid code that can be compiled successfully. However, compilation feedback remains valuable because it provides an execution-level verification step that can detect environment-specific issues, such as missing dependencies or unresolved imports, which may not be identified through documentation-based validation alone.


Overall, the ablation study confirms that our system benefits from the coordinated interaction of its components. The Analyzer improves task understanding, Documentation Retrieval provides reliable external library information, Code Generation produces the program based on the collected knowledge, while Validation and Compilation enable verification and correction. Together, these components form a complementary pipeline that progressively reduces library-related failures and improves the reliability of LLM-generated code.

\begin{tcolorbox}[arc=0pt, colback=gray!5, colframe=black!70, title=Answer to RQ2, top=2mm, bottom=2mm, right=2mm, left=2mm]
The ablation study confirms that each component contributes to the effectiveness of the proposed system. Removing any individual component decreases correctness and increases library-related errors, demonstrating their complementary roles in improving generated code reliability. The complete configuration achieves the best results, reaching 83\% correctness while minimizing library-related errors and affected files.
\end{tcolorbox}

\section{Discussion}
\label{sec:discussion}

In this section, we discuss the key findings of our study. We first interpret the experimental results, then examine the contribution of the proposed agentic design, assess its generalizability across different language models, and conclude with the broader implications of our work.

\subsection{Mitigation of Library-Related Errors}

The results presented in Section~\ref{sec:results1} show that library-related issues remain a persistent challenge for LLM-based code generation, even when using recent and capable models. Across all evaluated models, direct prompting leads to incorrect dependency usage, including invalid imports, deprecated APIs, and unsupported library components. These findings are consistent with prior studies showing that LLMs often struggle with API evolution and library misuse due to their reliance on pre-trained knowledge that may not reflect the current state of software ecosystems~\cite{zhuo2025identifyingmitigatingapimisuse,latendresse2025robustllm}. 

The improvements achieved by our approach demonstrate that supplementing LLMs with external and up-to-date library knowledge can effectively reduce these limitations. As shown in Section~\ref{sec:results1}, integrating documentation retrieval and validation reduces library-related errors across all evaluated models, indicating that these errors are not solely caused by insufficient model capability. Instead, they are partly caused by the absence of mechanisms that connect code generation with the current state of external libraries. By retrieving relevant documentation and validating generated code against library constraints, the proposed approach provides information that is complementary to the model's pre-trained knowledge.

However, our approach does not completely eliminate all library-related failures. Some errors remain when the retrieved documentation does not provide sufficient information about specific APIs, when multiple libraries interact in complex ways, or when the generated solution requires reasoning beyond the available documentation. These cases highlight opportunities for future research, such as improving documentation retrieval strategies, incorporating version-aware reasoning, and developing more advanced validation mechanisms.

Overall, our findings suggest that improving LLM-based code generation requires not only increasing model capabilities but also integrating external knowledge and verification mechanisms that can adapt to the continuously changing software ecosystem.




\subsection{Contribution of Individual Components}

The results of the ablation study presented in Section~\ref{sec:results2} demonstrate that the effectiveness of the proposed approach comes from the interaction between its specialized components rather than from a single module. Removing individual components increases library-related errors, showing that each stage addresses a distinct source of failure in the code generation process.

The impact of documentation retrieval highlights the importance of providing LLMs with current and task-specific API information. Without this component, the model must rely mainly on its pre-trained knowledge, which can contain outdated or conflicting library usage patterns. Similarly, the contribution of the Analyzer demonstrates the importance of explicitly identifying required dependencies before generation. By making dependencies explicit, the system reduces incorrect assumptions about library availability and usage. The contribution of the Compilation and Validator modules further confirms that generation alone is insufficient, as automatically generated code requires continuous verification to detect and correct issues that cannot be identified during generation.

These observations align with previous research on retrieval-augmented generation and agent-based software engineering systems, which shows that combining generation with external knowledge sources and feedback mechanisms can improve reliability in complex tasks~\cite{wang2025coderagbenchretrievalaugmentcode}. Our results extend these findings by showing that such mechanisms are particularly valuable for library-dependent code generation, where correctness depends not only on program logic but also on compatibility with evolving external APIs.

Nevertheless, the ablation results also reveal that no individual component completely solves library-related failures. The system achieves the best performance when dependency analysis, documentation grounding, generation, and validation operate together. This suggests that future approaches should focus on adaptive combinations of specialized components rather than relying exclusively on a single generation model.

\subsection{Generalization Across Language Models}

An important observation from our evaluation is that the proposed approach improves performance across diverse LLM families, including proprietary and open-weight models. The consistent reduction of library-related errors across GPT-5, DeepSeek-V3, Qwen3, Mistral, and Llama 3 suggests that the approach is not tightly coupled to a specific model architecture or training strategy.

The magnitude of improvement varies across models due to differences in their pre-trained knowledge and their ability to reason about external libraries. Each model has been trained on different data sources and knowledge cutoffs, leading to varying degrees of exposure to library APIs and software frameworks. Consequently, some models are more likely to rely on outdated API knowledge or generate hallucinated library usages. By grounding code generation in up-to-date documentation and validating the generated code against the current execution environment, the proposed approach complements the models' pre-trained knowledge with current library information. The consistent improvements observed across all evaluated models indicate that library-related errors are not specific to a particular LLM but stem from a common limitation of relying solely on pre-trained knowledge for code generation.


\subsection{Implications}
The findings of our study provide concrete actionable insights into the challenges and solutions surrounding library-aware code generation. In this section, we discuss the implications of our results for practitioners, researchers, and educators.

\subsubsection{Implications for Developers} 
Our results show that library-related errors remain common across all evaluated LLMs when code is generated using direct prompting alone. In particular, \textit{incorrect import paths} consistently represented the most frequent category of errors, indicating that current LLMs struggle to keep pace with rapidly evolving software libraries and changing package structures. These findings suggest that relying solely on an LLM's internal knowledge is insufficient for producing reliable, dependency-heavy code. Software teams should therefore integrate documentation-aware retrieval mechanisms into AI-assisted development workflows, allowing models to retrieve current library documentation before generating code \cite{wang2025coderagbenchretrievalaugmentcode}. By grounding generation in up-to-date documentation, practitioners can avoid the time-consuming process of manually scouring external release notes or API references to diagnose broken imports.

Our ablation study further demonstrates that documentation retrieval alone is insufficient for maximizing code correctness. While the Documentation Retrieval module produced the largest standalone improvement, removing the validation and refinement stages still resulted in lower execution success. This finding highlights the importance of combining documentation grounding with automated verification. Organizations developing AI-assisted software engineering tools should therefore integrate compilation, dependency validation, and iterative refinement into their development pipelines to detect and resolve library-related issues before deployment. Automating these validation and refinement steps eliminates tedious trial-and-error debugging loops, saving developers substantial time and engineering bandwidth that would otherwise be spent manually resolving subtle, cascading dependency failures \cite{fu2025autorepairtestcasesllms, shinn2023reflexionlanguageagentsverbal}.

Beyond improving correctness, our findings suggest that organizations should view documentation retrieval and automated validation as complementary components of AI-assisted development rather than optional enhancements. Integrating these capabilities into existing IDEs and continuous integration (CI) pipelines can help identify dependency-related issues early, reducing debugging effort while increasing developer confidence in LLM-generated code before it reaches production \cite{trooskens2026compiledaideterministiccode}.

\subsubsection{Implications for Researchers} Our findings demonstrate that accurate dependency identification is a fundamental component of library-aware code generation. The Analyzer agent, which constructs structured dependency specifications prior to documentation retrieval, played a key role in enabling the framework to retrieve relevant documentation while avoiding unnecessary context. These results suggest that future research should focus on intelligent context selection rather than simply expanding context windows with large volumes of documentation. Developing methods that automatically identify the libraries, APIs, and relevant documentation snippets required for a given programming task has the potential to further improve the reliability and efficiency of retrieval-augmented code generation systems \cite{ gu2025retrieveeffectiveretrievalaugmentedcode}.

Although our framework reduced library-related errors by up to 54.6\%, a portion of generated programs still failed after the maximum number of refinement iterations. This suggests that iterative validation alone is insufficient to recover from certain classes of dependency-related errors once generation diverges from a correct solution. Future work should therefore investigate more effective error recovery strategies, including richer compiler feedback, adaptive documentation retrieval, improved multi-agent collaboration, and fine-tuning approaches that leverage iterative debugging and code repair traces. Such techniques could further improve the robustness of agentic code generation systems beyond the capabilities demonstrated in this work \cite{wu2026runtimeexecutiontracesguided}.

Our results also highlight the need for standardized evaluation methodologies that explicitly measure library-related correctness rather than relying solely on traditional functional correctness benchmarks. Future benchmarks should include evolving software libraries, version-specific APIs, and dependency-resolution tasks to better reflect the challenges encountered in real-world AI-assisted software development.

\section{Threats to Validity}
\label{sec:threats}

In this section, we discuss threats to the validity of our work.

\subsection{Internal Validity}
Threats to internal validity involve the extent to which our findings accurately reflect the phenomena studied without being compromised by bias or confounding factors. The construction of the evaluation benchmark represents a potential source of bias. The prompts were initially generated with the assistance of an LLM based on real-world implementations collected from GitHub repositories involving rapidly evolving Python frameworks. While this approach facilitated the creation of realistic and diverse tasks, automated prompt generation may introduce variations in task formulation or inadvertently influence the difficulty of the resulting problems. To mitigate this threat, all generated prompts were manually reviewed, refined, and validated by the authors to ensure alignment with the original developer intent, preservation of the targeted library usage challenges, and consistency across benchmark instances.

Another potential threat concerns the manual identification and categorization of library-related errors in the exploratory study. Human judgment may introduce classification bias, particularly for borderline cases involving deprecated, ambiguous, or alternative API usages. To reduce this risk, two annotators independently analyzed all generated programs using predefined annotation guidelines. Disagreements were resolved through discussion, and the final annotations achieved a Cohen’s kappa coefficient of $\kappa = 0.86$, indicating strong agreement between annotators.

\subsection{External Validity}

Threats to external validity involve the extent to which our findings hold in other contexts. Threats to external validity may arise from the choice of evaluation frameworks, as the characteristics of the selected ecosystems may influence the observed performance of the proposed system. To mitigate this concern, we selected LangChain, AutoGen, and CrewAI, which are considered the most widely used frameworks for developing LLM-based applications. These ecosystems are characterized by rapid evolution and frequent API updates, making them highly representative and challenging environments for studying library-related errors in LLM-generated code. Nevertheless, the findings may not directly generalize to more stable libraries or ecosystems in other programming languages, such as Java, JavaScript, or C++. As part of future work, we plan to evaluate our system on a broader set of frameworks and programming languages to further assess its generalizability across diverse software ecosystems.

Another threat is the generalizability of our results to different underlying LLM. Different models possess varying baseline capabilities, context window sizes, and instruction-following accuracies, which could influence the absolute performance metrics. To mitigate this, we evaluated our approach across five distinct models of varying sizes and architectures, demonstrating that the relative improvements provided by our workflow are consistent regardless of the underlying base model.

\section{Related Work}
\label{sec:related}
In the following, we first discuss works related to characterizing failure patterns in LLM-generated programs, with a focus on library and API-related errors. We then review prior work related to improving code generation reliability through library-aware strategies, external knowledge retrieval, and API grounding.

\subsection{Errors in LLM-generated code}

Prior studies have shown that LLM-generated code suffers from systematic errors that limit its reliability in practical software engineering settings. Empirical investigations have identified recurring failure modes across models, prompts, and programming tasks, revealing limitations beyond traditional human-written bugs~\cite{wang2025understandingcharacteristicscodegeneration, tie2024llmsimperfectwhatempirical}.
Tambon et al.~\cite{tambon2024bugslargelanguagemodels} conducted a large-scale analysis of LLM-generated bugs and proposed a taxonomy including hallucinated objects, incorrect attributes, and semantic inconsistencies. Their findings show that LLM-generated errors differ from conventional software bugs, requiring specialized analysis approaches.

A growing body of work focuses on dependency and library-related hallucinations~\cite{krishna2025importingphantomsmeasuringllm, twist2025libraryhallucinationsllmsrisk}. These studies demonstrate that LLMs frequently generate non-existent APIs, deprecated methods, and incorrect imports. Such failures are particularly common in rapidly evolving frameworks, where outdated training knowledge leads models to reference obsolete or invalid components~\cite{10.1145/3728894}.
Spracklen et al.~\cite{spracklen2025packageyoucomprehensiveanalysis} further show that LLMs may recommend non-existent packages or misuse existing libraries through incorrect aliases, missing submodules, or incompatible API combinations. These issues directly affect code executability and may introduce security risks.
Despite these findings, existing evaluations of LLM-generated code largely emphasize functional correctness, often overlooking library- and import-related failures~\cite{liu2023codegeneratedchatgptreally}.

\subsection{Library-aware and API-grounded code generation}

A growing body of research focuses on improving library- and API-aware code generation to reduce errors in LLM outputs. These approaches aim to mitigate hallucinated dependencies, incorrect imports, and misused APIs by grounding generation in external knowledge and structuring the synthesis process. 
One line of work imposes library-aware generation strategies. For example, CodeGen4Libs~\cite{liu2023codegen4libs} generates import statements before the main code, reducing inconsistent or unused dependencies. A\textsuperscript{3}-CodGen~\cite{liao2024a3codgenrepositorylevelcodegeneration} enriches generation with repository-level context, including global project structure, local files, and third-party libraries, helping to avoid library incompatibilities and logical errors. While these methods improve consistency within a stable project, they are less effective for frameworks undergoing rapid structural reorganization where the "correct" repository-level context is a moving target.

Another strand emphasizes retrieval-augmented or API-grounded generation. De-Hallucinator~\cite{eghbali2024dehallucinatormitigatingllmhallucinations} iteratively retrieves relevant documentation to refine prompts and improve correct API usage. Related methods leverage formal API specifications or natural-language documentation at generation time, ensuring outputs align with verified interfaces rather than relying solely on the model’s internal knowledge~\cite{patil2023gorillalargelanguagemodel, li2023apibankcomprehensivebenchmarktoolaugmented}. Recent research also frames correct library usage as a planning and exploration problem. ExploraCoder~\cite{wang2025exploracoderadvancingcodegeneration} decomposes unfamiliar API usage into structured subtasks to guide correct invocation. Other agentic approaches integrate execution feedback, tool calls, or iterative self-refinement to detect and repair invalid imports or API calls~\cite{madaan2023selfrefineiterativerefinementselffeedback, peng2024perfcodegenimprovingperformancellm}. These methods tightly couple generation with validation, producing more reliable code in the presence of complex libraries and evolving dependencies. 

\section{Conclusion}
\label{sec:conclusion}

In this work, we investigated library-related issues in LLM-generated code and proposed an agentic approach to mitigate these errors. Our exploratory study of 100 LLM-generated programs revealed that 84\% of instances contain at least one library-related error, with common patterns including incorrect import paths, missing imports, hallucinated libraries, deprecated usage, and unused imports. Based on these findings, we designed a library-aware generation approach combining dependency analysis, documentation grounding, code generation, and automated validation. We evaluated our approach on 300 programming tasks from rapidly evolving Python frameworks across five LLMs. The results show that our system consistently improves code correctness, reducing library-related errors by 38.1\%--54.6\% and increasing correctness by up to 16\%. These findings demonstrate the effectiveness of documentation grounding and validation mechanisms for improving the reliability of LLM-generated code.

Future work will explore extending library-aware reasoning to broader software engineering tasks, such as API migration, dependency management, and repository-level development. We also plan to integrate additional software knowledge sources and execution feedback to further improve the adaptability of LLM-based agents.

\section*{Data Availability}
The datasets, implementation, and outputs used in this study are publicly available at~\cite{lagc_repo}.

\bibliographystyle{IEEEtran}
\bibliography{software}

\end{document}